\documentclass[aps,12pt,pra,showpacs,superscriptaddress,preprint]{revtex4}

\usepackage{amsmath}

\usepackage[dvipdfmx]{graphicx}

\usepackage{subfig}

\usepackage{array}

\usepackage{amsfonts}

\usepackage{epsf}

\usepackage{epsfig}

\usepackage{amssymb}

\usepackage{bbm}

\usepackage{delarray}

\usepackage{caption}

\usepackage{amstext}

\usepackage{graphics}

\usepackage{epstopdf}

\usepackage{xcolor}

\catcode`ð=\active

\defð{\u{g}}

\catcode`Ð=\active

\defÐ{\u{G}}

\catcode`Ý=\active

\defÝ{\. I}

\catcode`ö=\active

\defö{\"{o}}

\catcode`Ö=\active

\defÖ{\"O}

\catcode`ü=\active

\defü{\"{u}}

\catcode`Ü=\active

\defÜ{\"{U}}

\catcode`Þ=\active

\defÞ{\c{S}}

\catcode`þ=\active

\defþ{\c{s}}

\catcode`ý=\active

\defý{{\i}}

\catcode`ç=\active

\defç{\d{c}}

\catcode`Ç=\active

\defÇ{\d{C}}

\begin{document}

\title{Solutions of Pauli-Dirac Equation in terms of Laguerre Polynomials within Perturbative Scheme}
\author{\small Altuð Arda}
\email[E-mail: ]{arda@hacettepe.edu.tr}\affiliation{Division of
Physics Education, Hacettepe University, 06800, Ankara,Turkey}

\begin{abstract}
We search for first- and second-order corrections to the energy levels of the Pauli-Dirac equation within the Rayleigh-Schrödinger theory. We use some identities satisfied by the associated Laguerre polynomials to reach this aim. We give a list presenting analytical forms of some integrals including two associated Laguerre polynomials, or their derivatives.
\end{abstract}


\maketitle

\newpage

\section{Introduction}
The Pauli-Dirac equation is an extended form of the Dirac equation based on a non-minimal coupling, which describes the interaction due to magnetic moment of a spin-$\frac{1}{2}$ particle [1]. The particle could have an electric charge, or be neutral. The interaction of a neutral particle having an anomalous magnetic moment with the electromagnetic field is a particular situation, because a duality exists between the anomalous magnetic moment and the electric charge of electron [2, 3, 4], which is called the Ahanorov-Casher effect [5].

The exact solutions of the Pauli-Dirac equation for different electromagnetic fields (for a constant magnetic field and an electromagnetic plane wave) can be found in textbooks [6]. It is also possible to find some works about the solutions of the equation in literature, where the bounded solutions, and the corresponding wave functions of a neutral spin-$\frac{1}{2}$ particle for various electric fields, have been investigated by Lin [7], the solutions of the Pauli-Dirac equation for a neutral particle in an electromagnetic field have been presented as a "quasi-exact problem" by Ho, and co-workers [8], and the exact solutions of the Pauli-Dirac equation have been researched within the position-dependent mass formalism by Arda et. al. [9]. In the present work, we deal with the bounded solutions of the equation within the Rayleigh-Schrödinger perturbation theory, where also present analytical results of some integrals including two associated Laguerre polynomials, and their first-derivatives.

The Laguerre polynomials, and integrals including them with analytical and/or numerical results, have received a special interest within physics. They appear in normalization of the wave functions within non-relativistic/relativistic quantum mechanics in three, and $D$-dimensional cases [10, 11, 12]. The Laguerre polynomials are also strongly related with the information measures, and position- and momentum-space entropies (especially Fisher's, Renny's, and Shannon's entropy) [13], the spreading measures [14, 14], probability distributions [16] within the quantum information theory, and have a key importance in applied and computational mathematics and combinatorics [14]. The expectation values of position in non-relativistic and relativistic Coulomb potential in terms of the Hahn and Chebyshev polynomials, where integrals of the Laguerre polynomials are written from them, have been extensively researched in Ref. [17]. An extensive list has been given in Ref. [18], where the authors have studied QED corrections to bound state energies for the Coulomb problem. Mavromatis and co-workers have presented the analytical results of some particular integrals including the Laguerre polynomials from a purely mathematical point of view [19, 20].

The organization of the work as follows: In Section II, we briefly obtain four coupled differential equations for the Pauli-Dirac equation for a neutral spin-$\frac{1}{2}$ particle in spherical electric field. We present second-order differential equations for each of two radial wave functions. In Section III, we search the perturbative solutions up to second-order of the Pauli-Dirac equation by setting a particular electric field configuration within the Rayleigh-Schrödinger theory. We obtain the analytical results for the corrections coming from first- and second-order corrections by using some identities satisfied by the associated Laguerre polynomials. We give also some integrals having analytical results of two associated Laguerre polynomials, or their first derivatives in Appendix. The conclusions are given in last section.

\section{Pauli-Dirac Equation for Spherical Electric Field}

The Pauli-Dirac equation, which describes a neutral spin-$\frac{1}{2}$ particle with mass $M$, having a magnetic moment $\mu$, and moving in an external electromagnetic field, is given [1, 7-9]
\begin{eqnarray}
\left(i\gamma^{\mu}\partial_{\mu}-\frac{1}{2}\,\mu\sigma^{\mu\nu}F_{\mu\nu}-M\right)\Psi(\vec{r},t)=0\,,
\end{eqnarray}
where $F_{\mu\nu}$ is the field strength, $\gamma^{\mu}=(\gamma^{0}, \vec{\gamma\,})$ are the Dirac matrices, and $\sigma^{\mu\nu}=\frac{i}{2}[\gamma^{\mu}, \gamma^{\nu}]$. The field strength tensor is written in terms of external electric field $\vec{E}$ and magnetic field $\vec{B}$ as following
\begin{eqnarray}
\frac{1}{2}\,\sigma^{\mu\nu}F_{\mu\nu}=i\vec{\alpha}.\vec{E}-\vec{\Sigma}.\vec{B}\,,
\end{eqnarray}
where $\vec{\alpha}=\gamma^{0}\vec{\gamma}$, and $\Sigma^{k}=\frac{1}{2}\,\epsilon^{ijk}\sigma^{ij}$. For the case where external fields are time-independent, the four-component spinor $\Psi(\vec{r},t)$ can be decomposed into two parts as
\begin{eqnarray}
\Psi(\vec{r},t)=e^{-i\varepsilon t}\psi(\vec{r}\,)\,,
\end{eqnarray}
Inserting Eq. (3) into Eq. (2) gives the stationary Pauli-Dirac equation
\begin{eqnarray}
\left(\vec{\alpha}.\vec{p}+i\mu\vec{\gamma}.\vec{E}-\mu\beta\vec{\Sigma}.\vec{B}+\beta M\right)\psi(\vec{r}\,)=\varepsilon\psi(\vec{r}\,)\,,
\end{eqnarray}
where $\vec{p}=-i\nabla$ and $\beta=\gamma^{0}$.

In the rest of the present work, we focus only the case where the magnetic fields are absent. The stationary Pauli-Dirac equation then takes the form combined of four coupled partial differential equations
\begin{subequations}
\begin{align}
(\varepsilon+M)\psi_2=\vec{\sigma}.(\vec{p}-i\mu\vec{E})\psi_1\,,\\
(\varepsilon-M)\psi_1=\vec{\sigma}.(\vec{p}+i\mu\vec{E})\psi_2\,,
\end{align}
\end{subequations}
where we write $\psi=(\psi_1, \psi_2)^{t}$ ($t$ denotes transpose), and $\psi_1$ and $\psi_2$ are two-component spinors, respectively, with the following representation
\begin{eqnarray}
\vec{\alpha}=\begin{pmatrix}
0 & \vec{\sigma}\\
\vec{\sigma} & 0
\end{pmatrix}\,,\,\,\,\beta=\begin{pmatrix}
1 & 0 \\
0 & -1
\end{pmatrix}\,,\nonumber
\end{eqnarray}
where $\vec{\alpha}$ are the Pauli matrices.

For a spherical symmetric electric field $\vec{E}=E(r)\hat{r}$, where $\hat{r}$ is unit vector, one can describe a complete set of operators as $(H, \vec{J}^{2}, J_{z}, \vec{S}^{2}, K)$ because we deal with vanishing magnetic fields. The operator $H$ in this set is the Hamiltonian given in Eq. (4), $\vec{J}$ is the total angular momentum $\vec{J}=\vec{L}+\vec{S}$, where $\vec{L}$ is the orbital angular momentum, and $\vec{S}=\frac{1}{2}\,\vec{\Sigma}$ is the spin operator. The operator $K=\beta(1+\vec{\Sigma}.\vec{L})$ satisfies the commutation relations $[H, K]=[\vec{J}, K]=0$ [7-9].

Writing the first part of solution to Eq. (5) as $\psi^{+}=(\psi^{+}_1, \psi^{+}_2)^{t}$ with
\begin{eqnarray}
&&\psi^{+}_1(\vec{r}\,)=F_1(r)\varphi^{+}(\vec{r}\,)\,,\nonumber\\
&&\psi^{+}_2(\vec{r}\,)=iF_2(r)(\vec{\sigma}.\hat{r})\varphi^{+}(\vec{r}\,)\,,
\end{eqnarray}
gives us two first order differential equations for the radial wave functions
\begin{subequations}
\begin{align}
\left(\frac{d}{dr}+\mu E(r)-\frac{L}{r}\right)F_1(r)&=-(\varepsilon+M)F_2(r)\,,\\
\left(\frac{d}{dr}-\mu E(r)+\frac{L+2}{r}\right)F_2(r)&=(\varepsilon-M)F_1(r)\,,
\end{align}
\end{subequations}
where $\varphi^{+}(\vec{r}\,)$ is two-component spinor written in terms of spherical harmonics $Y_{L\,m}(\theta, \vartheta)$:
\begin{eqnarray}
\varphi_{L m}^{+}(\vec{r}\,)=\begin{pmatrix}
  \sqrt{\frac{L+m+1}{2L+1}}\,Y_{L m}(\theta, \vartheta)\\
  \sqrt{\frac{L-m+1}{2L+1}}\,Y_{L m}(\theta, \vartheta)
  \end{pmatrix}\,,
\end{eqnarray}
We have finally a second order differential equation for the radial wave function $F_1(r)$ as
\begin{eqnarray}
\left\{\frac{d^2}{dr^2}+\frac{2}{r}\,\frac{d}{dr}+\varepsilon^2-M^2+\mu\,\frac{dE(r)}{dr}-\mu^2E^2(r)+2(L+1)\mu\,\frac{E(r)}{r}-\frac{L(L+1)}{r^2}\right\}F_1(r)=0\,.\nonumber\\
\end{eqnarray}

The other part of solution to Eq. (5) as $\psi^{-}=(\psi^{-}_1, \psi^{-}_2)^{t}$, where
\begin{eqnarray}
&&\psi^{-}_1(\vec{r}\,)=F_3(r)\varphi^{-}(\vec{r}\,)\,,\nonumber\\
&&\psi^{-}_2(\vec{r}\,)=iF_4(r)(\vec{\sigma}.\hat{r})\varphi^{-}(\vec{r}\,)\,,
\end{eqnarray}
with $\varphi^{-}(\vec{r}\,)$ as two-component spinor written in terms of spherical harmonics $Y_{L\,m}(\theta, \vartheta)$:
\begin{eqnarray}
\varphi_{L m}^{-}(\vec{r}\,)=\begin{pmatrix}
  \sqrt{\frac{L-m}{2L+1}}\,Y_{L m}(\theta, \vartheta)\\
  -\sqrt{\frac{L+m}{2L+1}}\,Y_{L m}(\theta, \vartheta)
  \end{pmatrix}\,,
\end{eqnarray}
We obtain two coupled, first order differential equations for radial wave functions with the help of Eq. (10) as
\begin{subequations}
\begin{align}
\left(\frac{d}{dr}+\mu E(r)-\frac{L+2}{r}\right)F_3(r)&=-(\varepsilon+M)F_4(r)\,,\\
\left(\frac{d}{dr}-\mu E(r)-\frac{L}{r}\right)F_4(r)&=(\varepsilon-M)F_3(r)\,,
\end{align}
\end{subequations}
which give a second order differential equation for the radial wave function $F_4(r)$ as
\begin{eqnarray}
\left\{\frac{d^{2}}{dr^2}+\frac{2}{r}\,\frac{d}{dr}+\varepsilon^2-M^2-\mu\,\frac{dE(r)}{dr}-\mu^2E^2(r)-2(L+1)\mu\,\frac{E(r)}{r}-\frac{L(L+1)}{r^2}\right\}F_4(r)=0\,.\nonumber\\
\end{eqnarray}

We present analytical results for the energy levels of the Dirac-Pauli equation within the perturbation scheme in the next section. Because of similarity between Eqs. (9) and (13) we only deal with the perturbative solutions of Eq. (9) for which we set a particular electric field configuration.

\section{Perturbative Solutions}

To obtain the energy eigenvalues of the Dirac-Pauli equation perturbatively we need a spherically symmetric electric field configuration. A particular setting such as $E(r)=A+Br$ is suitable for this aim, and also makes it possible to compare the results with the known ones. Here, we indicate that this configuration can be expanded with a third term proportional to $1/r$, and the problem can be solved for this extended field configuration. We restrict ourselves in the present work that $A$, and $B$ real.

\subsection{Solutions for Unperturbed Hamiltonian}

By inserting the above electric field, and writing the wave function in Eq. (9) as $(F_1)_n(r)=f_n(r)/r$ we can separate the unperturbed Hamiltonian
\begin{eqnarray}
H_0=\frac{d^2}{dr^2}-a^2_1+\frac{a^2_2}{r}-\frac{L(L+1)}{r^2}\,,
\end{eqnarray}
and first- and second-order perturbations to the Hamiltonian $H_0$, respectively, as
\begin{subequations}
\begin{align}
H_1&=-2\mu^2ABr\,,\\
H_2&=-\mu^2A^2B^2r^2\,,
\end{align}
\end{subequations}
with
\begin{subequations}
\begin{align}
-a^2_1&=\varepsilon^2_n-M^2-\mu^2A^2+\mu B(2L+3)\,,\\
a^2_2&=2\mu A(L+1)\,,
\end{align}
\end{subequations}
The Hamiltonian combining of the defined ones above as $H=H_0+H_1+H_2$ satisfies Eq. (9), and the energy eigenvalue equation of the unperturbed Hamiltonian can be given by $H_0f_n=\varepsilon^{(0)}_nf_n$, which means that $f_n$'s correspond to the unperturbed wave functions denoting as $\psi_n^{(0)}$ in the standard language of the Rayleigh-Shcrödinger perturbation theory.

We search first the wave functions in the energy eigenvalue equation for $H_0$, which should be finite for $r \rightarrow \infty$, and $r \rightarrow 0$. Because of the boundary conditions we write as
\begin{eqnarray}
f_n(r) \sim r^{L+1}e^{-a_1 r}u_n(r)\,,
\end{eqnarray}
Inserting it into the energy eigenvalue equation for $H_0$ gives us
\begin{eqnarray}
x\,\frac{d^2u_n(x)}{dx^2}+(2L+2-x)\,\frac{du_n(x)}{dx}-\left[\frac{a^2_2}{2a_1}-(L+1)\right]u_n(x)=0\,,
\end{eqnarray}
which is a Kummer's-type (or confluent hypergeometric) equation with the solutions [21, 22]
\begin{eqnarray}
u_n(x) \sim \,_1F_1\left(\frac{a^2_2}{2a_1}-(L+1); 2L+2; x\right)\,.
\end{eqnarray}
where $\,_1F_1(a; b; x)$ the confluent hypergeometric functions of the first type [21, 22]. This solution satisfies only the condition for $r \rightarrow 0$, but it takes the form $e^{2a_1 r}$ for $r \rightarrow \infty$, so it should be cut in order to get a finite solutions as
\begin{eqnarray}
\frac{a^2_2}{2a_1}-(L+1)=-n\,;\,\,n=0, 1, 2, \ldots\,.
\end{eqnarray}
where $n$ corresponds to the radial quantum number. The last equation gives the unperturbed energy eigenvalues as
\begin{eqnarray}
\varepsilon^{(0)}_n=\sqrt{M(L)+\mu^2A^2\left[\frac{(n+L+1)^2-(L+1)^2}{(n+L+1)^2}\right]\,}\,\,;\,\,M(L)=M^2-\mu B(2L+3)\,,
\end{eqnarray}
and we write the unperturbed wave functions in terms of the associated Laguerre polynomials $L_{n}^{\alpha}(x)$
\begin{eqnarray}
f_n(x)=a_n\,\frac{n!(2L+1)!}{(n+2L+1)!}\,\left[\frac{n+L+1}{2\mu A(L+1)}\right]^{L+1}\,e^{-x/2}x^{L+1}L_{n}^{2L+1}(x)\,,
\end{eqnarray}
where the new variable $x$ is given as $r=\frac{n+L+1}{2\mu A(L+1)}\,x$, $a_n$ is the normalization constant, and used the equality $\,_1F_1(-n; \alpha+1; x)=\frac{n!\Gamma(\alpha+1)}{\Gamma(n+\alpha+1)}\,L_{n}^{\alpha}(x)$ [21, 22]. We need the wave functions satisfied Eq. (9) in the rest of the work, so we write
\begin{eqnarray}
(F_1)_{n}(x)=a_n\,\frac{n!(2L+1)!}{(n+2L+1)!}\,\left[\frac{n+L+1}{2\mu A(L+1)}\right]^{L}\,e^{-x/2}x^{L}L_{n}^{2L+1}(x)\,.
\end{eqnarray}
At this point, it is worth to say that the energy spectrum given in Eq. (21) holds only for particles ($+\varepsilon^{(0)}_n$), and it is possible to have anti-particle's spectrum too ($-\varepsilon^{(0)}_n$), and we consider only particle's spectrum to analyze the problem within the perturbative scheme.

To end this section, we obtain the normalization constant $a_n$ for which we write
\begin{eqnarray}
\int_{0}^{\infty}|(F_1)_n(r)|^2r^2dr+\int_{0}^{\infty}|(F_4)_n(r)|^2r^2dr=1\,,
\end{eqnarray}
We only use the first term to identify the normalization constant because the whole equality in Eq. (24) gives very complicated result. Instead of that one, we give not of all, but some of integrals containing of two associated Laguerre polynomials or their first-order derivatives, and having nonzero off-diagonal elements required for obtaining the normalization constant in Appendix. The normalization constant in Eq. (23) is given by
\begin{eqnarray}
a_n=\left[\frac{2\mu A(1+L)}{n+L+1}\right]^{L}\frac{1}{(n+L+1)^2(1+2L)!}\,\sqrt{\frac{4[\mu A(1+2L)]^3(n+2L+1)!}{n!}\,}\,,
\end{eqnarray}
and then the normalized, unperturbed wave functions are
\begin{eqnarray}
(F_1)_n(x)=Q(n,L)\,e^{-x/2}x^{L}L_{n}^{2L+1}(x)\,,
\end{eqnarray}
where
\begin{eqnarray}
Q(n,L)=\frac{2}{(n+L+1)^2}\sqrt{\frac{[\mu A(1+2L)]^3n!}{(n+2L+1)!}\,}\,.
\end{eqnarray}

\subsection{First and Second-Order Corrections to Energy Levels}

We find first- and second-order corrections to the energy spectrum of the Dirac-Pauli equation according to the definitions [23]
\begin{subequations}
\begin{align}
\varepsilon^{(1)}_n&=<n|H_1|n>\,,\\
\varepsilon^{(2)}_n&=<n|H_2|n>+\sum_{i \neq n}\frac{|<i|H_1|n>|^2}{\varepsilon^{(0)}_n-\varepsilon^{(0)}_i}\,,
\end{align}
\end{subequations}
where $|n>$ denotes the normalized, unperturbed wave functions $(F_1)_n(x)$. As the power of the independent variable increases details of the computation gets complicated, and the terms of the series are named according to the logic of perturbation theory.
The following equality of the associated Laguerre polynomials [21, 22]
\begin{eqnarray}
xL_n^{\alpha}=(2n+\alpha+1)L_n^{\alpha}-(n+1)L_{n+1}^{\alpha}-(n+\alpha)L_{n-1}^{\alpha}\,,
\end{eqnarray}
will help us to evaluate the diagonal and nondiagonal elements of the operators $H_1$ and $H_2$ given in Eq. (28).

With the help of Eqs. (26) and (27), and by using Eq. (29) in Eq. (28) we write
\begin{eqnarray}
x|n>=2(n+L+1)|n>-Q_1[n](n+1)|n+1>-Q_2[n](n+2L+1)|n-1>\,,
\end{eqnarray}
where
\begin{subequations}
\begin{align}
\frac{Q(n,L)}{Q(n+1,L)} \equiv Q_{1}[n]&=\left(\frac{n+L+2}{n+L+1}\right)^2\sqrt{\frac{n+2L+2}{n+1}\,}\,,\\
\frac{Q(n,L)}{Q(n-1,L)} \equiv Q_{2}[n]&=\left(\frac{n+L}{n+L+1}\right)^2\sqrt{\frac{n}{n+2L+1}\,}\,,
\end{align}
\end{subequations}
and obtain the first-order corrections to the energy levels as
\begin{eqnarray}
\varepsilon^{(1)}_n=-2\mu^2AB<n|r|n>=-\mu B\frac{n+L+1}{1+L}<n|x|n>=-2\mu B\frac{(n+L+1)^2}{1+L}\,.
\end{eqnarray}

By multiplying Eq. (30) with $x$ again, and following similar steps, we obtain the first term in Eq. (28b) including diagonal elements of the matrix operator as
\begin{eqnarray}
&<n|H_2|n>=-\mu^2B^2A^2<n|r^2|n>=-\left[\frac{B(n+L+1)}{2(1+L)}\right]^2<n|x^2|n>\nonumber\\
&=-\left[\frac{B(n+L+1)}{2(1+L)}\right]^2\left[4(n+L+1)^2+(n+1)(n+2L+2)+n(n+2L+1)\right]\,,
\end{eqnarray}
We also need the off-diagonal terms of the Hamilton operator $H_1$ in Eq. (28b), which can be written as
\begin{eqnarray}
\sum_{i \neq n}\frac{|<i|H_1|n>|^2}{\varepsilon^{(0)}_n-\varepsilon^{(0)}_i}&=&\frac{1}{\varepsilon^{(0)}_n-\varepsilon^{(0)}_{n+1}}\left(\frac{n+L+2}{n+L+1}\right)^4(n+2L+2)(n+1)\nonumber\\&+&
\frac{1}{\varepsilon^{(0)}_n-\varepsilon^{(0)}_{n-1}}\left(\frac{n+L}{n+L+1}\right)^4n(n+2L+1)\,.
\end{eqnarray}

With the help of Eq. (21), we are now able to write the whole result for the energy spectrum of the Dirac-Pauli equation including also the first- and second-order corrections
\begin{eqnarray}
\varepsilon_n&=&\varepsilon^{(0)}_n+\varepsilon^{(1)}_n+\varepsilon^{(2)}_n\nonumber\\
&=&\sqrt{M(L)+\mu^2A^2\left[\frac{(n+L+1)^2-(L+1)^2}{(n+L+1)^2}\right]\,}-2\mu B\frac{(n+L+1)^2}{1+L}\nonumber\\&-&\left[\frac{B(n+L+1)}{2(1+L)}\right]^2\left[4(n+L+1)^2+(n+1)(n+2L+2)+n(n+2L+1)\right]\nonumber\\
&+&\frac{\left(\frac{n+L+2}{n+L+1}\right)^4(n+1)(n+2L+2)}{\sqrt{M(L)+\mu^2A^2\frac{n^2+2n(1+L)}{(n+L+1)^2}\,}-\sqrt{M(L)+\mu^2A^2\frac{(n+1)^2+2(n+1)(1+L)}{(n+L+2)^2}\,}}\nonumber\\&+&
\frac{\left(\frac{n+L}{n+L+1}\right)^4n(n+2L+1)}{\sqrt{M(L)+\mu^2A^2\frac{n^2+2n(1+L)}{(n+L+1)^2}\,}-\sqrt{M(L)+\mu^2A^2\frac{(n-1)^2+2(n-1)(1+L)}{(n+L)^2}\,}}\,.
\end{eqnarray}
We observe that the second, and third terms are independent from the parameter $A$ while the first part of the perturbed Hamiltonian includes the parameter. The third term is proportional to square of $B$ although it is a correction one coming from the first-order.


\section{Conclusions}
We have searched the perturbative corrections up to second-order to energy levels of the Pauli-Dirac equation within the Rayleigh-Schrödinger theory by using some identities of the associated Laguerre polynomials. We have also presented a list of integrals of two associated Laguerre polynomials, and their first derivatives required for the normalization of the wave function. We have found out the results for a particular electric field having a spherical symmetry. We have observed that the first-order corrections are proportional with the parameter $B$, and square of it, and independet of the parameter $A$ although the Hamiltonian $H_1$ is proportional only with $A$, and $B$. The second, and third terms in Eq. (35) are interesting because they are independent from the parameter $A$ although it comes from first-order correction. All terms coming from second-order correction are dependent on $B$, and $A^2$, which are placed in the square root of denominator. The whole contributions coming from first- and second-order corrections are also dependent on both of the radial, and angular momentum quantum numbers $(n, L)$.


\section{Appendix}

We give a list of some integrals including associated Laguerre polynomials with the following generic forms
\begin{eqnarray}
I(\eta,\xi)=\int_{0}^{\infty}dxe^{-x}x^{2L+\eta}L_{n}^{2L+\xi}L_{i}^{2L+\xi}\,,\nonumber
\end{eqnarray}
and
\begin{eqnarray}
J(\eta,\xi)=\int_{0}^{\infty}dxe^{-x}x^{2L+\eta}L_{n}^{2L+\xi}\left(\frac{d}{dx}L_{i}^{2L+\xi}\right)\,,\nonumber
\end{eqnarray}
which are explicitly
\begin{eqnarray}
I(0,1)&=&\left\{
  \begin{array}{ll}
    -\frac{(n+2L+1)!}{(n+1)!} & \hbox{for i=n+1;} \\
    \frac{(n+2L+1)!}{n!}\frac{2n+2L}{(n+2L+1)(n+2L)} & \hbox{for i=n;} \\
    -\frac{(n+2L)!}{n!} & \hbox{for i=n-1.}
  \end{array}
\right.\nonumber\\
I(1,1)&=&\frac{(n+2L+1)!}{n!}\,,\nonumber\\
I(2,1)&=&\left\{
  \begin{array}{ll}
    -\frac{(n+2L+2)!}{n!} & \hbox{for i=n+1;} \\
    \frac{2(n+2L+1)!}{n!}(n+L+1) & \hbox{for i=n;} \\
    -\frac{(n+2L+1)!}{(n-1)!} & \hbox{for i=n-1.}
  \end{array}
\right.\nonumber\\
I(3,1)&=&\left\{
  \begin{array}{ll}
  \frac{(n+2L+1)!}{n!}(n+2L+2)(n+2L+3) & \hbox{for i=n+2;} \\
    -\frac{2(n+2L+1)!}{n!}(n+2L+2)(2n+2L+3) & \hbox{for i=n+1;} \\
    -\frac{2(n+2L+1)!}{n!}n(n+2L+1) & \hbox{for i=n;} \\
    \frac{2(n+2L+1)!}{n!}n(2n+2L+1) & \hbox{for i=n-1.}\\
    \frac{(n+2L+1)!}{n!}n(n-1) & \hbox{for i=n-2.}
  \end{array}
\right.\nonumber\\
J(1,1)&=&\left\{
  \begin{array}{ll}
    -\frac{(n+2L+1)!}{n!(n+1)}(2n+2L+1) & \hbox{for i=n+1;} \\
    \frac{(n+2L+1)!}{n!}\left[\frac{2n(n+L+1)}{(n+2L+1)(n+1)}+n+2L+2\right] & \hbox{for i=n;} \\
    -\frac{(n+2L+1)!}{n!}\frac{2n+2L+1}{n+2L+1} & \hbox{for i=n-1.}
  \end{array}
\right.\nonumber
\end{eqnarray}
\begin{eqnarray}
J(2,1)&=&\left\{
  \begin{array}{ll}
    0 & \hbox{for i=n+1;} \\
    \frac{n(n+2L+1)!}{n!} & \hbox{for i=n;} \\
    -\frac{n(n+2L+1)!}{n!} & \hbox{for i=n-1.}
  \end{array}
\right.\nonumber\\
J(3,1)&=&\left\{
  \begin{array}{ll}
    -\frac{(n+2L+1)!}{n!}n(n-1) & \hbox{for i=n+2;} \\
    -\frac{(n+2L+1)!}{n!}\left[\frac{n^2}{n+2L+1}+2n(n+L)\right] & \hbox{for i=n+1;} \\
    \frac{(n+2L+1)!}{n!}n(3n+4L+3) & \hbox{for i=n.}\\
    -\frac{(n+2L+1)!}{n!}n(n+2L+2) & \hbox{for i=n-1;}\\
    0 & \hbox{for i=n-2;}
  \end{array}
\right.\nonumber
\end{eqnarray}

\end{document}